\documentstyle[11pt]{article}
\newcommand{\s}{\sigma }                                              
\newcommand{\SI}{\Sigma }    
\newcommand{\tS}{\tilde{\Sigma}}
\newcommand{\tG}{\tilde G}    
\newcommand{\si}{\sigma _1}  
\newcommand{\st}{\sigma _2}

\newcommand{\xn}{x_{n}}
\newcommand{\xns}{x_{n}(\sigma )}

\newcommand{\xm}{x_{m}}

\newcommand{\e}{e^{i\int k_{0}Y}}                                      
\newcommand{\te}{e^{i\int k_{0}\tY _0}}                                  

\newcommand{\kim}{ k_{1}^{\mu}}                                      
\newcommand{\kom}{ k_{0}^{\mu}}                                      
\newcommand{\ki}{ k_{1}}
\newcommand{\yn}{ Y_{n}}                                             
\newcommand{\ym}{ Y_{m}} 
\newcommand{\kn}{ k_{n}}

\newcommand{\km}{ k_{m}}

\newcommand{\ko}{ k_{0}}

\newcommand{\kon}{ k_{0}^{\nu}}

\newcommand{\lpp}{\mbox {$e^{i\int _{c} \alpha (t) 
k(t) \partial _{z} X(z+t) dt +ik_{0}X}$}}

\newcommand{\gvk}{ e^{i\sum _{n }k_{n}Y_{n}}}                      
\newcommand{\gvks}{ e^{i\sum _{n\ge 0 }\int d\s k_{n}(\s )Y_{n}(\s )}} 
\newcommand{\GVKS}{e^{i\sum _{n \ge 0 }\int d\s ' K_{n}(\s ')\tY _{n}
(\s ')}}

\newcommand{\dsi}{\frac{\partial}{\partial x_{1}}}             
         
\newcommand{\dsis}{\frac{\partial}{\partial x_{1}(\sigma _{1})}} 
\newcommand{\dsts}{\frac{\partial}{\partial x_{1}(\sigma _{2})}}

\newcommand{\dsq}{\frac{\partial }{\partial x_{n+m}}}

\newcommand{\dds}{\frac{\delta}{\delta \sigma}}

\newcommand{\dsb}{\frac{\partial ^{2}[\tS +\tilde G] (\si ,\st )}
{\partial
 x_{1}(\si )\partial x _{1}(\st )}}
\newcommand{\dsnm}{\frac{\partial ^{2}}
{\partial x_{n}\partial x_{m}}}

\newcommand{\p}{\partial}                                           
\newcommand{\pp}{\partial ^{2}}                                     
\newcommand{\mup}{\partial _{\mu}}                                  
                                  
\newcommand{\eA}{\be    \label{lv}
:e^{i\{ \int d\s\ko (\s ) Y (\s )+ i\sum _{n>0}  \kn (\s ) \frac{\p
Y(\s )}{\xns}
 \} }:\]\[e^{ \int \int d\si
d\st \{ \ko (\si )\ko ( \st )[\tS +\tilde G](\si ,\st ) + (\sum _{n>0}
 \kn
(\si ) .
 \ko (\st ) \frac{\p
[\tS + \tilde G] (\si , \st )}{\p \xn (\si )}+
\si \leftrightarrow \st) \}} \] 
\[e^{\int \int d\si d \st \{ \sum _{n,m>0}\kn (\si ) .\km (\st )
\dsb  \}}
\ee}
\newcommand{\li}{ \lambda_{1}}                                    
                                    
\newcommand{\eps}{ \epsilon}                                        
\newcommand{\al}{\alpha }                                             
\newcommand{\aln}{\alpha _{n}} 
\newcommand{\tY}{\tilde Y}

\newcommand{\ai}{\mbox{$\alpha _{1}$}}

\newcommand{\at}{\mbox{$\alpha _{2}$}}

\newcommand{\la}{\mbox{$ \lambda $}} 
\newcommand{\be}{\begin{equation}}
\newcommand{\br}{\begin{eqnarray}}
\newcommand{\ee}{\end{equation}} 
\newcommand{\er}{\end{eqnarray}}

\renewcommand{\theequation}{\thesubsection.\arabic{equation}}

\begin{document} 
\title{
\hfill\parbox{4cm}{\normalsize IMSC/2000/02/07\\
                               hep-th/0008005}\\        
\vspace{2cm}
Loop Variables and Gauge Invariant Interactions - II
\thanks{This is a detailed description
of an
 approach, outlined in a talk at the
Puri Workshop
in 1996, to use loop variables to describe string interactions.}}
\author{B. Sathiapalan\\ {\em Institute of Mathematical Sciences}\\
{\em Taramani                     
}\\{\em Chennai, India 600113}}                                     
\maketitle 

\begin{abstract} 
 We continue the discussion of our previous paper on writing down gauge 
 invariant interacting equations for a bosonic string using the loop 
variable 
 approach.
 In the earlier paper the equations were written down 
in one higher dimension where the fields are
 massless. In this paper we describe a procedure for dimensional 
reduction that
 gives interacting equations for fields with the same spectrum as 
in bosonic 
string theory. We also argue that the on-shell scattering amplitudes 
implied 
by these equations for the physical modes are the same as for the 
bosonic 
string. We check this explicitly for some of the simpler equations.
The gauge transformation of space-time fields induced by gauge 
transformations of the loop variables are discussed in some detail.
The unintegrated (i.e. before the Koba-Nielsen integration), 
regularized version of the equations, are gauge invariant off-shell
(i.e. off the {\em free} mass shell). 

\end{abstract}
\newpage      

\section{Introduction} 

In an earlier paper \cite{BS0} (hereafter `I') we had described a method 
of writing down                                                        
gauge invariant interacting 
equations of motion for the modes of the bosonic string 
using the loop variable approach \cite{BS1,BS2,BS3,Puri}. 
As in the free case the 
equations were written down in one higher dimension where all 
the modes are 
massless. Interactions were introduced by the simple prescription
of thickening the string to a ``band''.

In the free case the dimensional reduction can easily be done, 
leading to 
equations for massive modes \cite{BS1}. The masses are essentially put
 in by
assigning some fixed value to the momentum ``$p_5$'' in the internal
 direction 
and choosing $(p_5)^2=m^2$ where $m$ is the rquired mass. While the 
procedure 
is adhoc and does not admit any simple geometric interpretation,
it neverthelss 
has the advantage that it provides a simple method of writing down 
gauge invariant
equations for the massive modes of the string.

When we consider interactions, this dimensional reduction 
 is not so straightforward and it remained to be demonstrated that some 
generalization of this method would work. This is done in this paper. We 
describe
a dimensional reduction procedure that gives the same free spectrum and 
also 
show that 
the scattering amplitudes that can be deduced from 
the nonlinear terms in the
equations of motion are consistent with the scattering amplitudes of 
string theory for on-shell 
physical states.
The dimensional reduction procedure is also 
consistent with gauge invariance.
Furthermore, as in the higher dimensional situation,
 gauge invariance does not 
require that the fields be on-shell. The gauge transformations and gauge
invariance are easy to describe when the Koba-Nielsen integrals have
not been performed, and there is a regulator on the world sheet.
We thus have a gauge invariant
set of equations that are valid off the (free) mass shell, 
i.e where the fields are not forced to satisfy $p^2=m^2$.
This is essentially what string field theory gives \cite{BP,SZ,W}.
 We should hasten to add that
 unlike string field theory we do not have an action that gives the 
equation of motion. It remains an open problem to find such an action. 
After the Koba-Nielsen integrals are done and the continuum
limit on the world sheet taken, we get a ``low energy'' effective
equation of motion. The issue of gauge invariance of these equations
is subtle. A preliminary calculation was done in I, but we do not 
pursue this in this paper. This issue is analogous to the 
question in string field theory where the higher order non-linearity 
of the both the equations and gauge transformations can emerge
only after integrating out degrees of freedom.

Apart from the fact that it is always a good idea to have different 
ways of
 solving a given problem, we believe that a utility of the present 
approach 
is that the gauge transformations have a simple form when expressed 
in terms
 of loop variables. In fact, gauge invariance is manifest in these 
variables and 
the problem is to show that a consistent set of gauge transformations 
can be {\em defined} for the space time fields.

 The gauge transformations in the free case look
 like local (along the string) scale transformations \cite{BS1}. 
This continues to be so even in the interacting 
case and also after dimensional reduction. 
However this is not at all manifest when we work with space time fields.
It suggests that working directly with loop variables rather than 
with space 
time fields might provide some 
understanding
of the underlying principles of string theory.

This paper is organized as follows. In Sec 2 we describe the dimensional 
reduction
procedure. We also show that it does not affect arguments for gauge 
invariance given in
I. In Sec 3 we give an explicit calculation of 
some three and four point functions in 
the case of the tachyon and massless vector.

 In Sec 4 we conclude with a summary and future directions. 

\section{Dimensional Reduction}
\subsection{Reproducing String Amplitudes}

Let us first state our requirements as a motivation for the dimensional
 reduction prescription given below. We emphasize that our attitude 
here is 
that ultimately what we want are {\em space - time} 
gauge invariant (or coordinate invariant
- for closed strings)
 equations 
for the modes of the string with the caveat that the on shell S-matrix 
elements implied by these equations should be the same
 as that given by conventional string theory. Any set of rules
that achieves this is alright as long as they are self consistent. In 
particular we do not worry about {\em world sheet} reparametrizations 
or BRST
invariance or any such
elegant geometrical property. If they exist 
that would be a bonus, but we do
not demand any such interpretation at this stage.

 Thus, the dimensional reduction
 procedure
thus should not violate 
the gauge invariance that is built into the loop variable approach.
At the free level we simply required that $(p_5)^2 = m^2$ where $p_5$ is 
the momentum in the internal direction and $m$ is the mass of the field,
which is also related to the naive dimension `P' (more precisely,
 $m^2 = {P- 1\over \al '}$)
 of the vertex operator. We rewrite this
as $(p_5)^2 g^{55}=m^2$ to emphasize that there is a metric that 
could play
a role. When interactions are taken into account one expects that
$p_5$ of the different interacting fields will add up as required by
 momentum
conservation. It would seem that all values of $p_5$ have to be allowed.
But we do not want this.
Because  only the string oscillators in the higher dimension,
and not the zero mode, are expected
to contribute space-time degrees of freedom 
in the usual covariant string field constructions \cite{SZ}.
In order to retain this feature we will set $p_5 =1$ for all fields. 
We will
further require that when there are interactions 
\be     \label{g55}
g^{55}={P-1 \over N^2}
\ee

where $N$ is the number of fields at the interaction. Thus $N=1$ for 
the free 
case and $N=2$ for the quadratic term in the equation of motion, etc.
The net effect of this is that $(p_5 + q_5 + k_5 ...)^2 g^{55} = P-1$ is 
 true for every term in the equation of motion.
This is a very peculiar looking ansatz without any obvious
geometrical interpretation.
\footnote{ There are other ways of achieving these ends, that
might make it look a little more geometrical - for instance
one can modify the range of the $\s$  integration in
 $\int d\s p_5 X^5(\s )$
relative to the $\s $ integration for the other directions.
For all directions (other than 5) we let the range of $\s$ integration
be $0$ to  $ N$. For the $5$ direction we assume it is $0$ to $1$ always.
This brings in a relative factor of $N^2$ that we need.} 
Nevertheless it serves the purpose of providing us the massive equation
starting from the massless equation in one higher dimension. 

 As will be seen below, this 
is crucial in deriving the equation of motion. We 
will also set $X^5=0$ at the end in 
order that there be no momentum conservation in the internal direction.

In the correlation functions we have terms of the type 
$...(z-w)^{p.q}...$.
On expanding the dot product we get $p^\mu q_\mu + p^5q_5$. 
Again we do not 
want the $p^5q_5$ terms if we want to recover the Veneziano amplitudes.
One way to achieve this is to set, 
in the definition of the space time fields,
 \be    \label{p5q5}
< k_n(\s _i)k_m(\s _j) ....> = ...S_n S_m (z_i-z_j)^{-p_i^5 p_{j5}}....
\ee
 This will ensure that all unwanted factors of $(z-w)^{p^5q_5}$ 
are canceled. We will come back to this later.

These set of rules will be applied below in deriving the equations 
of motion.

Let us see why these rules reproduce the Veneziano amplitudes. 
The argument
is very similar to that made in \cite{BSPT}. For an N-point amplitude
there are N-3 Koba-Nielsen variables that need to be integrated. 
Three of them
can be fixed. In our case the N-point amplitude gives rise to a term
in the equation of motion with N-1 fields and therefore N-1 Koba Nielsen 
variables. One of these is trivial because of translational invariance. 
Thus we have N-2 integrations. The last (i.e. (N-2)th)integration 
will actually just produce (when the particles are on-shell)
a term of the form ${1\over (p_1 +p_2 +p_3 +...p_{N-1})^2-m^2}$.
(We have used $(p_{1,5}+p_{2,5}+...p_{N-1,5})^2 g^{55}= P-1 =m^2$.)
 This is just
the propagator for the external Nth field, whose equation of motion we 
are writing down, and whose mass is $m^2$. When we vary w.r.t the 
Liouville
mode we bring down a factor of $(p_1+p_2+...p_{N-1})^2-m^2$ which 
precisely
cancels this propagator. The remaining N-3 integrals then produce the 
Veneziano amplitude. 
As explained in \cite{BSPT} if we regularize the
integral we end up subtracting the intermediate poles in the amplitude. 
This 
is also exactly what needs to be done when constructing an effective 
action.

Note that varying the Liouville mode in terms involving its derivative,
 also
contributes terms to the equation of motion, but these are essentially
gauge covariantizations of the previous term. These terms are not there
for on shell physical states. Thus from this we conclude that for the 
on-shell
physical states this calculation give you the correct result. The rest
 of the 
terms are fixed by gauge invariance.   

Note that P was set equal to the naive dimension of the vertex operator. 
We have been
using $\kn Y_n$
\footnote{See Appendix A for the basic definitions}
 as our vertex operator. But $Y_n$ does not have a well
defined dimension. It contains all $\tY _n$ in it. Furthermore, even
$\tY$ doesn't have an unambiguous dimension because of the $z$ 
dependence.
Thus $\tY _n (z) = \tY _n (w) + (z-w) (n) \tY _{n+1} +...$. Both these
ambiguities have to be removed. The first will be removed by using $K_m$
defined in I by 
\be
\sum _n \kn \yn = \sum _m K _m \tY _m
\ee
This gives: ($\al _0 =1$)
\be    \label{2.02}
K_n = k_n + \al _1 k_{n-1} + \al _2 k_{n-2} +...+ \al _n \ko
\ee

They have the same gauge transformation law as $\kn$.

The second ambiguity will be removed by translating all vertex operators
to a canonical location on the world sheet that we call $z$.
Thus 
\be
\sum _n K_n \tY _n (w) = \sum _m K_m(z-w) \tY _m (z)
\ee

which gives:
\be    \label{2.04}
K_q(z) = \sum _{n=0}^{n=q}K_n (0)D_n^q (-1)^{q-n}z^{q-n}
\ee
where
\br
D_n^q &  = & ^{q-1}C_{n-1},\; \; n,q\ge 1 \nonumber \\
      &  = & {1\over q}, \; \; n=0 \nonumber \\
      &  = & 1 ,   \;\;           n=q=0
\er

Now $P \tY _n (z) = n \tY _n (z) $ is an unambiguous equation.

The gauge transformation law for $K_n (0)$ is the same as for $\kn$:
\be
K_n \rightarrow K_n + \la _p K_{n-p} = K_n + \la _p {d\over dx_p}K_n
\ee
Of course the second form of the gauge transformation rule cannot be 
written for $\kn$ because they are not functions of $\al _n$.
The gauge transformation of $K_n(z)$ however,
 can only be written in this form:

 \be \label{2.06}
K_n(z) \rightarrow K_n (z) + \la _p {d\over dx_p}K_n(z)
\ee
In view of the above
it is worth examining afresh the proof of gauge invariance given in I.

Our starting point was the loop variable $e^A$ given by:

\eA

The only change we make is that $e^B \equiv \gvk$ is rewritten as 
$e^B \equiv e^{i\sum _n K_n (z-w)\tY _n (z) }$. The change in B is
$\la _p {d\over dx_p}B$ by eqn (\ref{2.06}), which is just what it 
was earlier.
Also the operator ${P-1\over N^2}$ commutes with ${d\over dx_n}$, because
P acts only on $\tY _n$ which has no $\xn$ dependence.  

Thus the  crucial point in the proof, which was that   
\be
\delta A = \int d\s \li (\s )[\dsis + \dsts ] A 
\ee
is not affected.

It was also shown in I, that while gauge invariance at the level
of loop variables is almost manifest, what is non-trivial is that
one can consistently define gauge transformation rules for 
{\em space-time
fields} , i.e., the consistency of the {\em map} 
from loop variables and their transformations
to fields and their transformation is a non-trivial issue.
In I we mapped each loop variable expression of the form
$ \kn (\si ) \km (\st ) \tG (\si , \st )....$ to a space time 
field or fields
$S_n S_m \tG (\si , \st ) + S_{n,m} \tG (\si ,\si ) ...$ \footnote{
 We also kept 
the cutoff dependent second term which would normally be discarded 
during the normal ordering process}. This whole procedure has to be 
reexamined
for the following reason: When we substitute $g^{55}= {P-1\over N^2}$,
 the
numerical value depends on which equation we are considering. Thus the
 same
term, say, $S_{55}g^{55}$ occurring in an equation for the tachyon will
have a value of ${-1\over N^2}$ whereas in an equation for a 
massive particle
could be ${1\over N^2}$. The question is are we going to get mutually 
inconsistent transformation rules for the field $S_{55}$?
We describe below a way out of these problems.

\subsection{Consistent Definition of Gauge Transformation}
\subsubsection{Before Dimensional Reduction}

Let us begin with a general discussion of the consistency issue.
Let us consider a generic term 
\be   \label{2.0.10}
\kn (\s _i) \yn (z _i) \km (\s _j) \ym (z _j ) ....
\ee
We have suppressed the Lorentz indices. Under a gauge transformation
it transforms to 
\be    \label{2.0.11}
\int d\s \la _p (\s )[k_{n-p} (\s _i) \yn (\s _i,z_i) \km (\s _j) 
\ym (\s _j,z_j ) + ....]
\ee

The reason for retaining the $ \yn $ in the above is that when one maps
the above gauge transformation to space time fields one has to be 
careful   
about factors of the form $\tG (z_i -z_j)$ that arise out of the
 contraction
of the $\yn 's$. The resulting $z_i-z_j$ dependence will, after doing the
Koba Nielsen integrals (over $z_i, z_j$) introduce some non trivial
momentum dependence in the gauge transformation laws. These are the 
factors
$F(p,q)$ in eqn.(6.0.9) and (6.0.14) of I.

 We will first go over to the $K_n$ variables of (\ref{2.02}).
One can then transfer the $z_i$ dependence into the
$K_n $ and use $\tY (z)$ with some canonical $z$ as in (\ref{2.04}).
The point $z$ is arbitrary.
We will remove this arbitrariness by defining it to be the location
of $\la (\s )$ , i.e. $z(\s )=z$ 

Thus we replace the above with:
\[
K_n (\s _i,z_i -z) \tY _n (z) K_m (\s _j,z_j-z) \tY _m (z) 
\rightarrow K_n (\s _i,z_i -z) \tY _n (z) K_m (\s _j,z_j-z) \tY _m (z)
\]
\be  \label{GTR}
+
\int d \s \la _p(\s ) {d\over dx_p}(K_n (\s _i,z_i-z) K_m (\s _j,z_j-z)
\tY _n(z)\tY _m (z))
\ee

When we contract the $\tY$'s we will have to do some regularization.
 We also have to introduce vertex operators $\e$ for the momentum
dependence. We will worry about each of these issues in turn. But first
let us see how gauge transformation laws for space time fields 
can be 
consistently defined from the above.
The first step is to substitute (\ref{2.02}) and (\ref{2.04})
 in the above
as the space time fields are defined in terms of $\kn$ and not $K_n$.

The leading term in (powers of $\al _n$ and $z_i-z$) is 
$\kn (\s _i )\km (\s _j )$ which
leads to $S_{n,m}$ (and $S_n  S_m$). All other space time fields have 
smaller quantum numbers $n,m$.
We get something that looks like 
\be   \label{GTR1}
\delta S_{n,m} = \Lambda _p S_{n-p,m} + \al _1 \Lambda _p S_{n-p-1,m} +
 ...
\ee
where we have ignored the powers of $z_i-z_j$. 
 Thus if we assume that we have already defined
a gauge transformation for all the lower quantum number space-time 
fields, then
(\ref{GTR}) defines uniquely a gauge transformation law for $S_{n,m}$. 
Of course the powers of Koba-Nielsen variables $z_i-z$ will translate to 
appropriate dependences on momenta when the integrals are done.
 For the moment
let us leave the dependences on $z_i$ as they are. 
Let us turn to the issue of contractions.
We can contract the $\tY _n$'s after point splitting or some other
 regularization. A simple regularization is to use $ln ((w-z)^2+\eps ^2)$
as the Green function. 
Thus in (\ref{GTR}) we will replace $\tY _n(z) \tY _m (z)$ by
$\tY _n (z) \tY _m (z')$ where $z'=z+\epsilon$ or use the regularized
Green function. In either case contraction will produce
 $\approx \eps ^{-n-m}$.
 We will get an
equation for the traces, $S_{n,m \mu}^\mu$, 
$S_{n\mu}S_m^\mu$, etc., which will
have an overall power of $(\epsilon )^{-n-m}$
 multiplying the whole equation,
(instead of $\tY _n ^\mu \tY _m ^\nu$).  

Suppose we had gone back one step and kept the $z_i$ dependence in the 
$\tY $'s.  
Then, instead of (\ref{GTR}) we would have:
\[
K_n (0) \tY _n (z_i) K_m (0) \tY _m (z_j) 
\rightarrow K_n (0) \tY _n (z_i) K_m (0) \tY _m (z_j)+
\]
\be  \label{2.0.14}
+
\la _p {d\over dx_p}(K_n (0) K_m (0)\tY _n(z_i)\tY _m (z_j))
\ee
 
Contracting the $\tY _n$'s now give derivatives of $G(z_i-z_j)$. 
If we now
expand $G(z_i-z_j)$ in powers of $z-z_i$ and $z -z_j$( where we have
regularized as before) and 
keep track of powers of $\epsilon$ we will recover the contracted 
version of
 (\ref{GTR}). 
We are guaranteed this because of the following simple fact:
\[
G(z_i-z_j)= <\tY _n(z_i) \tY _m(z_j)>=<\tY _n(z + z_i-z)
 \tY _m(z +z_j-z)>
\]
\[
=<\tY _n(z) \tY _m(z)> + (z_i-z)<n \tY _{n+1}(z) \tY _m(z)> + ....
\]
This is exactly the equation that fixes the
 $z$ dependence of $K_n$ and gives
\be
G(z_i-z_j) \approx {(n+m-1)!\over (n-1)!(m-1)!} (\epsilon )^{-n-m} + 
(z_i-z)n{(n+m)!\over n!(m-1)!}
(\epsilon )^{-n-m-1}+...
\ee
More generally we can define $\eps _{n,m}= <\tY _n (0) \tY _m (0)>$ and
write the Taylor expansion in terms of them. 
For the regularized Green function
introduced above , viz $ln ((z-w)^2 +\eps ^2)$ we can see that 
$\eps_{n,m}=\eps_{m,n}$. 
If we insert this in (\ref{2.0.14}) we reproduce 
the contracted version of
 (\ref{GTR}).

Two comments are in order before we proceed. As stated in I, the fields
$S_{n,m}$ must have an $\xn$-dependence because the RHS of the
 gauge transformation (\ref{GTR1}) has $\xn$ dependence. This can be done
as in I by making the string field, $\Psi$, $\xn$- dependent. 
Second, the 
grouping of terms used in defining gauge transformation for 
space time fields
is different from that used in I. The one used here is manifestly
 consistent
so we will use this from now on. The method used in I is only correct to
lowest order as it compares terms involving $V(z_i) V' (z_j)$ with
$V (z_i) V' (z_i)$ (for some vertex operators $V,V'$), 
which can only be true to lowest order.
There is also one very important point: Here we are dealing with
(regularized) operator products. But in an actual equation of 
motion the locations of the operators have to be integrated over.
Furthermore we may also be interested in taking $\eps \rightarrow 0$.
 Verifying the gauge invariance of the final equation starting
from the gauge transformations defined here may be quite a complicated
process. In this paper we will not attempt this.

Now we turn to the second issue of introducing the vertex operators $\e$.
This is very important when we discuss gauge transformations because
when the $\s$ in $\la _p (\s )$ is not equal to any of the $\s '$ in
$\kn (\s ')$ then we have an extra vertex operator
 $e^{i\ko (\s )X(z(\s ))}$
associated with $\la (\s )$. (In the contracted version this 
introduces extra factors
of $(z(\s )-z(\s '))^{\ko (\s ).\ko (\s ')}$.)  Also when we Taylor
expand $e^{i\ko (\s )X(z(\s ))}$ we introduce an infinite number
of new vertex operators. This latter fact in particular implies that
we cannot define the gauge transformation of an object such as
 (\ref{2.0.10})
We need to include {\em all} the vertex operators on both sides i.e. in
(\ref{2.0.10}) and in (\ref{2.0.11}) for consistency. 
Thus we have to define 
gauge transformation for the entire interacting loop variable $\gvks$. 
Thus we define
\be   \label{gt1}
\delta [\GVKS] = \int d\s d\si \la _p (\s) K _{n-p}(\si )\tY _n(\si )
e^{i\sum _{n\ge 0 }\int d\s ' K_{n}(\s ')\tY _{n}(\s ')}
\ee
 
If we allow contractions we just have to keep track of $\eps _{n,m}$
on both sides. This does not produce anything different. We will use
 the uncontracted version to define gauge transformations 
for simplicity.

We will assume that $z(\s )=z=0$ for convenience. We also have the
following relations:
\[
\sum _{q}K_q (-w) \tY _q (0) = \sum _n K_n (0)\tY _n (w)
\]
which defines $K_q(-w)$ to be given by
\be
K_q(-w)=\sum _{n=0}^q K_n (0) D_n^q (w)^{q-n}
\ee
where the $D_n^q$ have been defined earlier.
Thus, for instance
\[
K_1(-w)=K_1 + k_0 w
\]
\[
K_2(-w)=K_2 + K_1 w +K_0 {w^2\over 2}
\]
\[
K_3(-w)=K_3 + 2K_2 w +K_1 w^2 + K_0{w^3\over 3}
\]

Thus in (\ref{gt1}) we can specify more clearly the dependence
on $z(\s _i)$ by writing it as

\[
\delta [e^{i\int d\s \sum _n K_n (\s , -z(\s ))\tY_m (0)}]
\]
\be   \label{gt2}
= \int d\si \sum _n \int d\s \la _p (\s ){d\over dx_p}
[K_n(\si , -z(\si ))\tY_n(0)]
e^{i\int d\s ' \sum _m K_m (\s ', -z(\s '))\tY_m (0)}
\ee

\vskip 5mm
{\bf Vector -$ \mathbf A_1^\mu$}
\vskip 5mm

For the vector we compare coefficients of $\tY_1(0)$ on both sides.
This gives
\[
\delta [\int d\si K_1(\si , -z(\si ))\tY_1(0)
e^{i\int \ko (\s ') \tY_0(0)}]
\]
\[
= \int d\s \la _1(\s ) \int d\si \ko (\si ) \tY_1(0) 
e^{i\int \ko (\s ') \tY_0(0)}
\]
Thus
\[
\delta [\int d\si k_1(\si )\tY_1(0)
e^{i\int \ko (\s ') \tY_0(0)}]
\]
\[
=\int d\s \la _1(\s ) \int d\si \ko (\si ) \tY_1(0) 
e^{i\int \ko (\s ') \tY_0(0)}
\]
Taking the expectation value on both sides gives
\be
\delta [A_1^\mu (p)\tY_1^\mu (0)
e^{ip\tY_0(0)}]
= \Lambda (p)p^\mu \tY_1^\mu (0)
e^{ip\tY_0(0)}
\ee

\vskip 5mm
{ $ \mathbf S_2^\mu$}
\vskip 5mm

\[
\delta [\int d\si K_2(\si , -z(\si ))\tY_2(0)
e^{i\int \ko (\s ') \tY_0(0)}]
\]
\[
=\delta [\int d\si (K_2(\si)+ K_1(\si ) z(\si ) + 
\ko (\si ) {z(\si ) ^2\over 2})\tY_2
(0)
e^{i\int \ko (\s ') \tY_0(0)}]
\]

\[
= \int d\s \la _1(\s ) \int d\si [K_1(\si )+z(\si )\ko (\si )]
 \tY_2(0) e^{i\int \ko (\s ') \tY_0(0)}
\]

In terms of fields

\[
\delta [ S_2^\mu (p)e^{i p \tY_0(0)} \tY_2(0) 
+ \int d\si A_1^\mu (p) z(\si ) \tY_2(0)
e^{ip \tY_0(0)}]
\]
\[
=
\Lambda _{1,1}^\mu (p) \tY_2(0)e^{ip \tY_0(0)} + 
\Lambda _1(p) A_1^\mu (q) \tY_2(0)
e^{i(p+q) \tY_0(0)} +
\]
\be   \label{S2}
 \int d\s p^\mu \Lambda _1(p) z(\s )\tY_2(0) e^{ip \tY_0(0)}
\ee

The last term in the RHS of (\ref{S2}) is clearly the variation of the 
second term in the LHS. Thus we are left with

\[
\delta  S_2^\mu (k)e^{i k \tY_0(0)} \tY_2(0) 
=
\Lambda _{1,1}^\mu (k) \tY_2(0)e^{ik \tY_0(0)} + 
\Lambda _1(p) A_1^\mu (q) \tY_2(0)
e^{i(p+q) \tY_0(0)} 
\]

with the understanding that $p+q=k$.

Note that there is no $z$-dependence in the transformation law.
It is the same as the naive transformation law introduced
in \cite{BS2}.

\vskip 5mm
{ $ \mathbf S_{1,1}$}
\vskip 5mm

Let us compare the coefficients of $\tY_1(0)\tY_1(0)$.
\[
\delta [ {1\over 2!}\int d\si \int d\st K_1^\mu (\si ,-z(\si ))
 K_1^\nu (\st , -z(\st ))
\tY_1^\mu (0)\tY_1^\nu (0)
 e^{i\int \ko (\s ') \tY_0(0)}]
\]
\[
=
{1\over 2!} \int d\si \int d\st \int d\s \la _1(\s )
[\kom (\si ) (K_1^\nu (\st ) +\kon (\st )z(\st ))
+
\]
\[
\kon (\st ) (K_1^\mu (\si ) +\kom (\si )z(\si ))]
 e^{i\int \ko (\s ') \tY_0(0)}]
\]

On taking expectation values to bring in space-time fields, the LHS gives
\[
\delta  {1\over 2!}\int d\si \int d\st [S_{1,1}^{\mu \nu}
{D(\si -\st )\over a}
e^{i\ko (\si )\tY_0(0)} +
A_1^\mu (p)A_1^\nu (q)
e^{i(p+q)\tY_0(0)}
\]
\[
 +
A_1^\mu (p){D(\si -\st )\over a}p^\nu z(\st ) e^{ip\tY(0)}+
A_1^\nu (p){D(\si -\st )\over a}p^\mu z(\st ) e^{ip\tY(0)}]
\]

\[
=
{1\over 2}\delta [S_{1,1}^{\mu \nu} (k)e^{ik\tY} + 
A_1^\mu (p)A_1^\nu (q)e^{i(p+q)\tY}
+ \int d\si z(\si ) p^{(\nu }A_1^{\mu )}(p)e^{ip\tY}]
\]

RHS
\[
{1\over 2}\int d\si \int d\st \int d\s 
[{D(\s -\st )\over a}
{D(\s -\si )\over a}\Lambda _{1,1}^\nu (p) p^{\mu} 
+
\]
\[
{D(\s -\st )\over a}
\Lambda _1(p)p^{\mu}A_1^\nu (q)
+
{D(\si -\st )\over a}
\Lambda _1(p)q^\mu A_1^\nu (q)]+
\]
\[
2\int d\st \Lambda _1 (p) p^\mu p^\nu z(\st )
\]

Notice that the $z$-dependent terms are the same on the LHS and RHS
 we get

\be   \label{S11}
\delta S_{1,1}^{\mu \nu} = \Lambda _{1,1}^{(\mu}p^{\nu)} +
 \Lambda (p) q^{(\mu }
A_1^{\nu )}(q)s
\ee

Once again there are no $z$-dependent terms in the transformation law -
 it is the same
as the one defined in \cite{BS2}.

\vskip 5mm
{ $ \mathbf S_{2,1}^{\mu\nu}$}
\vskip 5mm

One can go through the same analysis for $\tY_2(0)\tY_1(0)$. One finds
terms that are $z$-independent, linearly dependent and quadratically
dependent on $z$. Using (\ref{S2}),(\ref{S11}) 
we find that the $z$-dependent
terms cancel and the net result is,

\be    \label{S21}
\delta S_{2,1}^{\mu\nu}= \Lambda _{1,1,1}^{\mu\nu} + 
\Lambda _{1,1}^\mu A_1^\nu +
\Lambda _1 S_{1,1}^{\mu\nu} + \Lambda _{2,1}^\mu (p) p^\nu + 
\Lambda _1(p)S_2^\mu (q)q^\nu
\ee

It is very interesting that the z-dependence cancels out in each case.
This is in contrast to the
gauge transformation defined in I. The point being that the vertex
operators being compared there were different.
Thus for instance $Y_2(z) = Y_2(0) + a zY_3(0) +bz^2 Y_4(0)+...$ (for 
some $a,b..$).
This implies that when we write down equations of motion in one basis,
it is a $z$-dependent linear combination of the equations in the other
basis.

In the basis used in this paper the $z$-dependence cancels out of the 
gauge transformation laws. They are the naive gauge transformation
 laws first 
introduced in \cite{BS2}. To see the invariance of the equations of I,
 one would have
to combine all the equations for the vertex operators
 $Y_1(z)$,$Y_2(z)$... 
and extract from them the coefficients of, say, $\tY _n(0)$.
 One should also
keep track of factors of $\eps_{n,m}$ - thus we might consider
the coefficient of say, $\tY _n(0)\eps_{p,q}$ This has to  be invariant
under the gauge transformations defined in this paper. It is actually 
quite easy to
see why this must be true. We know that the full equation defined by
$\dds \GVKS =0$ is invariant. Thus if we split up this equation into
various pieces labelled by $\tY _n (0) \eps _{p,q}..$ 
they must individually
vanish. However as mentioned earlier the situation {\em after}
the Koba-Nielsen variables are integrated is more subtle and needs 
further
study. As mentioned in the introduction, this is analogous to the situation
in string field theory, where gauge invariance of the string field 
cannot be directly
used to describe gauge invariance of the ``low energy'' equations of motion.
One has to solve for all the fields that have been integrated out.

 We show next that the dimensional reduction 
procedure
given before is consistent with this definition.

\subsubsection{After Dimensional Reduction}

The prescription for dimensional reduction is to set $g^{55}= 
{P-1\over N^2}$,
and also to set internal momenta $k_{05}$ for all fields to 1.
One can also introduce a ``vielbein'' $e^{5V}= \sqrt{P-1\over N^2}$
 so that
$e^{5V} e^5_V=g^{55}$. This will prove necessary in what follows.

We can separate (\ref{GTR}) into different sets of equations,
 depending on the 
number of $\tY ^5$'s. By Lorentz covariance these equations are 
independently
satisfied. In equations with more than one $\tY ^5$ one can contract them
to get a factor of $g^{55}$ for each pair of $\tY ^5$. Clearly these are
overall multiplicative factors in any equation.
Actually $P-1$ is an overall multiplicative factor and will not affect
the gauge transformation law.
 However $N$ can depend on the
particular term being considered. Thus for instance in the `55' equation
if the LHS has $S_{55}$ then $N=1$ and if the
RHS has $S_5S_5$ then $N=2$. Thus the `55' eqn is not obtainable 
from the `$\mu \nu $' equation by Lorentz covariance. This by itself 
is not a problem
of course, since we don't expect Lorentz invariance in the
 $5-\mu$ directions.
However when we consider the vertex operator $e^{ik_{05} \tY _0^5}$
 and consider contractions involving $X^5$ we encounter a problem.
Consider an equation that has the following form
\be    \label{2.2.21}
\delta [k_{n\mu}(\si ) k_{m5}(\st )] \tY _n^\mu \tY _m^5 \te
= \lambda _p (\s )  k_{n-p \mu}(\si )  k_{m5}(\st )\tY _n^{\mu}
\tY _m^5 \te
 +...         
\ee  

Note that there is no momentum conservation in the `5' direction. 
Also from now on to be consistent 
we use upper Lorentz indices on the $\tY _n $ and 
lower
indices on $\kn$.
The three 
dots refer to various terms involving higher powers of $z_i-z_j$
 and also 
other terms.
If we now contract $\tY _0 ^5$ with $\tY _n^ 5$
using $<\tY _m^5 \tY _0 ^5> \approx g^{55}\epsilon ^{-m}$
 we get the following:
\be     \label{2.2.22}
\delta [k_{n\mu}k_{m5}k_{05}g^{55}] \tY _n^\mu e^{ik_{05} \tY _0^5 }
\epsilon ^{-m}
= \lambda _p k_{n-p\mu}k_{m5}\ko ^5 \tY _n^\mu
  \epsilon ^{-m} +...         
\ee
We have suppressed the $\s$ arguments and
 written $k_{05} g^{55}= k_0^5$ on the RHS.
If we use the naive extension of 
\be
<k_{n \mu}>=S_{n\mu} \, \, <k_{n\mu}k_{m\nu}>= S_{nm \mu \nu}
\ee
to $\mu =5$ we
\be
<k_{n\mu}k_{m5}> = S_{n,m \mu 5} 
\ee
etc and also $k_{05}=1$.

This gives
\be    \label{2.27}
\delta S_{n,m \mu 5}\underbrace{g^{55}}_{P-1}\underbrace{k_{05}}_{1}
= \Lambda _{p+m}(p)S_{n-p \mu }(q)\underbrace{g^{55}}_{{P-1\over4}}
\underbrace{(p_{05}+q_{05})}_{2}
+...
\ee

We have used on the LHS
$k_{05} g^{55} =P-1$ and on the RHS $(p_0 +q_0)_5g^{55} ={P-1\over 2}$
 (because $g^{55}= {P-1\over N^2}$)
.

Whereas earlier without any contractions,
\be
\delta [k_{n\mu}(\si ) k_{m5}(\st )] 
= \lambda _p (\s )  k_{n-p \mu}(\si )  k_{m5}(\st )
+...
\ee
gave 
\be
\delta S_{nm \mu 5}=\Lambda _{m+p V}S_{n-p \mu} +...
\ee
which is not consistent with (\ref{2.27}).

 Thus we have a contradiction between the two 
equations. The source of the contradiction is the factor of $N$ in the 
definition of $g^{55}$. While we are free at the level of loop variables
to multiply by $g^{55}$ when we convert to space time fields
we cannot expect meaningful results if we use the naive expression
to define space time fields when $\mu=5$. In other words $g^{55}$
is not just a number since it keeps track of how many different
 fields there 
are.

A resolution of this problem is obtained by using the vielbein 
to convert 
everything to a flat `V' index.  This means  
a different definition of space-
time fields that have an index in the 5-direction. Define
\[
<k_{m5} e^{5V}> = <\km ^V >= S_m^V
\]
\be
<\kn ^\mu k_{m5} e^{5V}> = <\kn ^\mu \km ^V>= S_{n,m}^{\mu V}
\ee

Write (\ref{2.2.21})
\be    
\delta [k_{n\mu}(\si ) k_{mV}(\st )] \tY _n^\mu \tY _m^V e^{ik_{0V}
 \tY _0^V }
= \lambda _p (\s )  k_{n-p \mu}(\si )  k_{mV}(\st )\tY _n^{\mu}
\tY _m^V e^{i(p_{0V} +q_{0V} )
\tY _0 ^V } +...         
\ee
On contraction of the $Y$'s gives:
\be
\delta [k_{n\mu}k_{mV}] \tY _n^\mu k_0^V 
\epsilon ^{-m}
= \lambda _p k_{n-p\mu}k_{mV}(p_0 ^V+q_0 ^V) \tY _n^\mu
  \epsilon ^{-m} +...,  
\ee
where $\ko ^V = k_{05}e^{5V}$.
Converting to space time fields
\be
\delta S_{nm \mu V}=\Lambda _{m+p V}S_{n-p \mu}+...
\ee

 On the LHS  $\ko ^V$ has the value of $\sqrt {P-1}$.
On the RHS $e^5_V = {\sqrt {P-1}\over 2}$, so $p_0^V + q_0^V =\sqrt 
{P-1}$.
 Thus there is no contradiction.

In equations involving two or more $\tY_n ^5$ we do the same thing, 
namely
go to flat indices. So the equations take the form
\be
\delta S_{nmVV}\tY _n ^V \tY _m ^V
 =\Lambda _{n+p V}S_{m-pV}\tY _n ^V \tY _m ^V
\ee
Contraction of the $\tY ^V$'s does not introduce any factors of $g^{55}$ 
and thus there is no problem. The upshot of this discussion is that by
 going to flat coordinates the
`N' dependence is completely absent and there is thus no source of 
inconsistency from such contractions.

We have to now turn to the extra requirement (\ref{p5q5}).  This 
requirement
is effectively saying that we should include the factor 
$e^{i\int d\s k_{0,V} \tY_0^V(z(\s ))}$
in the definition of say $S_{11}$ as:
\[
<k_{1\mu} (\si )k_{1 \nu} (\st )e^{i\int d\s k_{0,V} \tY_0^V(z(\s ))}>
\tY _1^\mu (0)
\tY_1^\nu (0)
\]
\be   \label{newS11}
= [S_{11\mu \nu}{D(\si -\st )\over a} + S_{1\mu} S_{1\nu}]\tY _1^\mu(0)
\tY_1^\nu (0)
\ee

Note that the location of the exponential factor (which is {\em not} 
normal
ordered)
is at $z(\s )$ rather
than $0$. One can immediately see that 
when $\mu$ is replaced by $V$ the resulting contractions will induce
inconsistencies of the type mentioned above, to avoid which we went to 
great 
lengths. Thus the only solution is to have no possibility of
contraction between $\tY_0^V$'s. Thus we set
\be    \label{G55}
<\tY_0^V(z) \tY_0^V(w)> \equiv G^{VV} (z,w) =0
\ee

Since gauge transformations do not mix Lorentz indices, we are perfectly
free to choose this. Thus our dimensionally reduced loop variable
will take the form:
\[ 
exp \{
\int d\si \int d\st 
\sum _{n,m\ge 0}(\kn (\si ) .\km (\st ) {\p ^2 [\tG +\tS ] (\si ,\st )
 \over \p \xn (\si )\p \xm (\st )}
+ 
\]
\be   \label{newLV}
 k_{n, V} (\si )k_{m,V} (\st ) 
{\p ^2\tS  (\si ,\st ) \over \p \xn (\si )\p \xm (\st )}) \}
:exp \{ i\int d\s \sum _{n\ge 0}\kn \yn (z(\s )\}:.
\ee

In this case we do not need to modify our definition of space-time
fields as in (\ref{newS11}), nor do we need (\ref{p5q5}). We will
use (\ref{newLV}) as our loop variable.

We thus conclude that the dimensional reduction procedure does not 
introduce any inconsistency
in the definition of gauge transformation of space-time fields. Thus 
there
is a consistent map from loop variables to space time fields. Since the
equations were manifestly gauge invariant in terms of loop variables, we 
conclude that they are gauge invariant even when written in terms of
space-time fields. 

\section {Examples}
\subsection{Tachyon}

We begin with a discussion of the tachyon in order to illustrate the
dimensional reduction procedure.

\subsubsection{Preliminaries}

Let us fix some conventions first:
\be
<X^{\mu}(z_1) X^{\nu}(z_2)>=-2g^{\mu \nu} \, ln \, (z_1-z_2)
\ee

with $g_{00}=-1 \, ; g_{ii}=+1$\footnote{The regularized propagator
is $-g^{\mu \nu}ln ((z_1-z_2)^2 + \eps ^2)$.}.
 The mass shell condition in this convention
is $p^2=-m^2  $ and the tachyon has $m^2>0$. Using these conventions
\[
<e^{ik.X(z_1)} \, e^{-ik.X(z_2)}>= (z_1-z_2)^{-2k^2}
\]
From this it is clear that $e^{ik.X}$ has mass dimension $k^2$, and
furthermore that $\int dz e^{ik.X(z)}$ is a marginal operator when 
$k^2=1$.
This fixes $m^2=-1$ for the (open bosonic string) tachyon.
 
Another way to see this is to note that
\be   \label{NO}
e^{ik.X(z)}= e^{-{1\over 2}k^\mu k^\nu <X^\mu (z) X^\nu (z) >}
:e^{ik.X(z)}:=
\eps ^{k^2}:e^{ik.X(z)}:
\ee
where we have normal ordered the vertex operator and introduced a lattice
spacing $\eps$ as the ultraviolet cutoff. 
Scaling $\eps\rightarrow \la \eps$ scales the
ultraviolet momentum cutoff by $1\over \la$ or equivalently 
in units of the
cutoff increases the momentum of the operator by $\la$. This shows 
that the
mass dimension is $k^2$. 

Let us set $\eps =e^{2\s}$. If we change $k^2 $ to $k^2+ k_5^2$, then 
(\ref{NO}) becomes 
\be
e^{ik.X(z)}=e^{(k^2+k_5^2)\s}:e^{ik.X(z)}:
\ee
Imposing
$\dds e^{ik.X(z)}=0$ gives the equation 
\[
k^2+k_5^2=0
\]
Thus we let $k_5^2=-1$ we get the tachyon mass shell condition. In the
 general
case we need $k_5^2=( P-1)$ where $P$ is 
the engineering dimension of the vertex operator.

\subsubsection{Loop Variables}

Let us work out  the tachyon equation using the loop variable approach.
We start with
\[
e^{\int  d \si \int _0^1 d\st \ko (\si ).\ko (\st )[\tG (\si ,\st ) + 
\tS (\si , \st )]}
\]
\be   \label{TLV}
e^{\int  d \si \int _0^1 d\st k_{0V} (\si )k_{0V} (\st )
[\tS ^{VV}(\si , \st )]}
:e^{\int d\s \ko (\s )Y(\s ) +I(\s )}:
\ee

As explained in \cite{BS1} the tachyon vertex operator can be represented
by $I(\s )$.

{\bf Free Equation}

 For the free equation we have $0<\si <1$ and $\si = \st$.
 and we bring down only one power of $I$.
Thus we can set $\ko (\si )=\ko (\st ) = \ko$.
Using $\tG (\si , \si )\approx \eps$ , we get
\be
\int {dz\over \eps}I(\s )e^{(\ko .\ko + k_V k^V)\SI}
 \eps ^{\ko .\ko }e^{i\ko . \tY _0}
\ee  
where we have set all the $\xn$ to zero. Using $<I(\s )>= \Phi (\ko )$
we get on applying ${d\over d \SI}$ the equation
\be
\int dz \eps ^{\ko .\ko -1 }(\ko .\ko + k_V k^V)\Phi (\ko )=0
\ee

Further since $k_{0V}k_0^V = k_{05}k_{05}g^{55}=P-1=-1$ we finally get
\be
(\ko ^2 -1 )\Phi (\ko )=0
\ee

{\bf Cubic Interaction}

We let 
\[
0<\si <1 : \ko (\si )=p : z(\si )=z
\]
\[
1<\si <2 : \ko (\si )=q : z(\si )=w
\]

$ \int _0^2 d \si \int _0^2 d\st \ko (\si ).\ko (\st )[\tG (\si ,\st ) + 
\tS (\si , \st )]$ becomes 
\[
\{ p.p [\tG (\si ,\si )+ \tS (\si ,\si )] + p_Vp_V\tS (\si ,\si )+
 q.q [\tG (\st ,\st )+ \tS (\st ,\st )]+ q_Vq_V\tS (\st ,\st  )+
\]
\[
2p.q [\tG (\si ,\st )+ \tS (\si ,\st )]+  2p_Vq_V\tS (\si ,\st )\}
\] 

(\ref{TLV}) becomes
\[
{I(\si )I(\st )\over 2!}
e^{\{ p.p [G (\si ,\si )+ \tS (\si ,\si )] +  p_Vp_V\tS (\si ,\si )+
 q.q[G (\st ,\st )+ \tS (\st ,\st )]\}+ q_Vq_V\tS (\st ,\st )}
\]
\be
e^{[(2p.q + 2p_Vq^V) \tS (\si ,\st )]}
(z-w)^{2p.q}e^{i(p \tY _0(\si ) + q \tY _0(\st ))}
\ee

We have set $\xn=0$ and hence written $G$ and $\tY$. $G(z,z)=ln \, \eps$.
$\tS$ has to be Taylor expanded, but for the present we can just use the
lowest order approximation and replace them all by $\s (z)$. 
$<I(\si )I(\st )>= \Phi (p) \Phi (q)$. 
Also $g_{55}={-1\over 4}$.
Using all this we get to lowest order,
\be   \label{TLV1}
e^{[(p^2-1/4)+(q^2-1/4)+(2p.q-2/4)]\s}\int {dz\over \eps}
 \int {dw\over \eps} ({z-w\over \eps})^{2p.q}
e^{i(p+q)\tY _0}\Phi (p) \Phi (q) \eps ^{(p+q)^2-1}
\ee

The integral over $w$ is from $z+\eps$ to $\infty$. At infinity we assume
the contribution is zero. \footnote{Actually one should put an infrared
cutoff. In the proper time formalism \cite{BSPT} this is automatic. 
Here it has to be done by hand. If we are
going off shell this is important. In this paper we will ignore 
this issue
by assuming that we are always close to being on shell.} The 
integral gives,
at the lower end ${1\over 2p.q+1}$. When $p^2=q^2=1$, we can 
replace this by
$1\over(p+q)^2-1$. Applying $\dds$ to (\ref{TLV1})
gives $(p+q)^2-1$ which exactly cancels this. 
So we get
\be
\int dp \int dq {[(p+q)^2-1]\over 2p.q+1}\Phi (p) \Phi (q) e^{i(p+q)X} 
\approx \Phi ^2 (X)
\ee
The conclusion is that we get
for the cubic interaction between three on shell tachyons a momentum 
independent constant as expected from bosonic string theory.

{\bf Quartic Interaction}

Analogous to the previous calculation we introduce
\[
0<\s <1 : \ko (\s )=p : z(\s )=z_1
\]
\[
1<\s <2 : \ko (\s )=q : z(\s )=z_2
\]
\[
2 <\s <3 : \ko (\s )=k : z(\s )=z_3
\]
Furthermore we have,
\[
<I(\si )I(\st )I(\s _3)>\approx \Phi (p) \Phi (q)\Phi (k)
\]

Also as before $G(\s ,\s ) \approx \eps$ and $\tS \approx \s$.
$g_{55}= {-1\over 9}$. This gives
\[
e^{(p^2-1/9)+(q^2 -1/9)+(k^2-1/9) +(2p.q-2/9)+(2p.k-2/9)+(2k.q-2/9)]\s}
\]
\[
\int {dz_1\over \eps} \int {dz_2\over \eps} \int {dz_3\over \eps}
({z_1-z_2\over \eps})^{2p.q}({z_2-z_3\over \eps})^{2q.k}
({z_3-z_1\over \eps})^{2p.k}
\]
\be
\eps^{(p+q+k)^2}\Phi (p)\Phi (q) \Phi (k)e^{i(p+q+k)X}
\ee

The integrals can be evaluated in the usual way. First we use translation
invariance to set $z_3=0$. Then define $z_2'={z_2\over z_1}$ to get
\be
\int {dz_1\over \eps}
({z_1\over \eps})^{2p.k+2q.k+2p.q+1}\int dz_2' 
(1-z_2')^{2p.q}(z_2')^{2q.k}
\ee

The integral over $z_1$ gives $1\over 2p.q+2q.k+2k.p+2$. 
The action of $\dds$
gives $(p+q+k)^2-1$. These two factors cancel on shell. The resultant
integral over $z_2'$ is nothing but the usual Veneziano amplitude.
As shown in \cite{BSPT} regularizing the integrals by point splitting
subtracts the poles corresponding to on shell intermediate states.
 This gives the effective action.

\subsection{Vector}
For the massless vector $g^{55}=0$ and so it is fairly obvious that 
the pole
structure comes out right given the arguments given in the first section.
The only point of the calculation below is to illustrate the use of the
$K_n \tY _n$ variables as compared to $\kn \yn$ used in I.

We start with the loop variable
\[
e^{\int d \si \int d\st \ko (\si ).\ko (\st )[\tG (\si ,\st ) + 
\tS (\si , \st )]} 
\] 
\[
e^ {\int d \si \int d\st \{ \ki (\si ).\ko (\st ) \dsis
 [\tG (\si ,\st ) + \tS (\si , \st )] +\si \leftrightarrow \st \}}
\]
\be   \label{VLV}
:e^{i\int d\s [\ko (\s )\tY _0(z ) + K_1 (\s ,z-z (\s ))\tY _1 (z)]}:.
\ee

where $K_1(\s ,z-z(\s ))= \ki (\s ) + \ko (\s ) \al _1 +
 \ko (\s )(z-z(\s ))$.
 
\subsubsection{Free Theory}

For the free theory there is only one point.
\[
0<\s <1 : \ko (\s )=\ko , \ki(\s ) = \ki ; z(\s ) = z . 
\]

$G(\s ,\s ) \approx ln \, \eps $. In evaluating the derivative of the 
Green function one has to be more careful. Thus before we set
$\si = \st = \s$ we must either do point splitting
or use the regularized Green function $ln ((z-w)^2+\eps ^2)$.

Thus  let $\s _A$ and $\s _B$ be the
split points with $z(\s _{A,B})=z_{A,B}$. Of course $\ko (\s _A)=
\ko (\s _B)
=\ko$ and the same for $\ki$.
 
Then we have two possibilities

\[
\si = \s _A ,\, and \,  \st = \s _B  
\]
or
\[
\si = \s _B , \, and \, \st = \s _A
\]

We refer to $\xn (\si )$ by $\xn$ and $\xn (\st )$ as $y_n$.

Thus we have 
\[
\ki (\si ).\ko (\st ){d\over dx_1}G(\si ,\st , x,y)=
\ki (\s _A ).\ko (\s _B ){d\over dx_1}G(\s _A ,\s _B, x,y)
=\ki .\ko {1\over z_A-z_B}
\]
When we consider the second possibility we get
\[
\ki (\si ).\ko (\st ){d\over dx_1}G(\si ,\st ,x,y)=
\ki (\s _B ).\ko (\s _A ){d\over dx_1}G(\s _B ,\s _A, x,y)
=\ki .\ko {1\over z_B-z_A}
\]

Adding the two gives zero. This is because of the antisymmetric property
of the derivative of the Green function. 

If we use  $ln ((z-w)^2 +\eps ^2)$ it is easy to see that the 
derivative vanishes when $z \rightarrow w$. Thus using 
 either method we get the 
same result.

The derivative of $\tS$ does 
not have any subtlety and we simply get 
\[
\ki .\ko \dsi \SI (\s , \s , x,y) = \ki .\ko {1\over 2}\dsi 
\SI (\s , \s , x,x).
\]
We have used the fact that $\SI$ is symmetric in its arguments. 
Thus to lowest
order in $\xn$  we simply get
\[
\ki .\ko \dsi \SI 
\]
Thus (\ref{VLV}) becomes for the term proportional to $\tY _1$ 
(For this term $g_{55}=0$),
\[
\eps ^{\ko ^2} e^{\ko ^2 \SI } \ki .\ko
\dsi \SI e^{i(\ko \tY _0 + K_1 \tY _1)}+
\]
\be
\eps ^{\ko ^2} e^{\ko ^2 \SI } iK_1 \tY _1 e^{i(\ko \tY _0)}
\ee

When we vary w.r.t. $\SI$ we get (using $\dsi K_1 = \ko$)
\be
[-i\ko (\ki .\ko ) + \ko ^2 iK_1 ]\tY _1 e^{i\ko . \tY _0}=0
\ee
Using $k_{05}k_{05}g^{55}=0$ and $<\kim >=A ^\mu$, and setting $\xn =0$,
we recover Maxwell's equation
\[
\mup F^{\mu \nu}=0
\]

\subsubsection{Cubic term}

The first correction comes in at cubic order. 
The calculation is done in the Appendix
as an illustration. The terms that have a tachyon pole is specific to
the bosonic string. The other terms correspond to superstrings (and also
to Yang-Mills). One can compare these terms with known results
as given for instance in \cite{GSW} and see that they agree.

Thus we see in these examples 
 that on shell S-matrix elements are correctly reproduced
by this theory. General arguments were 
given in an earlier section.
 While this does not constitute a proof, hopefully that can also be
done with some more work.

In this paper,
we have not attempted to calculate higher order terms using the 
Taylor expansion of I. 
These
higher order contributions (coming from derivatives of $\SI$) 
to the equations of motion are additions 
that are dictated by gauge invariance. The point that needs to be 
stressed
is that any term that comes from {\em derivatives} of $\SI$ are 
``longitudinal'' or `` gauge'' pieces that can be set to zero if we
are only interested in S-matrix elements of physical degrees of freedom.
Thus they do not affect the arguments presented in this section,
 which were
to show that the S-matrix elements are reproduced correctly. Nevertheless
these terms are necessary when one wants manifest gauge invariance.

\section{Conclusions}

We have discussed how one can use loop variables to derive
gauge invariant equations of motion for all the modes
of the bosonic string. The first step was taken in I.
There the theory was written in one higher dimension where the
modes are all massless.
In this paper we have extended the results of I by explaining
how, by a particular type of dimensional reduction of the massless
theory,  we get the right spectrum and on shell scattering amplitudes
of bosonic string theory.  The theory is gauge invariant even off shell.
Thus in principle we have an off shell gauge invariant formulation.
We have not attempted a  proof of this result to all orders
and for all modes. However we have given some explicit examples 
and also arguments on why the method should work in general.

What the Koba-Nielsen integration does to gauge transformation
and equation, in   the continuum limit  has not been discussed
in this paper. This is an important question that deserves
further study.
Another issue that needs further work is the evaluation of higher
order corrections arising from the Taylor expansion of $\tS$.
We also need to work out the dimensionally reduced versions
of the equations of motion for massive particles (``$S_2$'' and 
``$S_{11}$''
of I) where one has to worry about the degrees of freedom
corresponding to $\kn ^5$, (they were called $q_n$ in \cite{BS1}).
It would also be interesting to get some geometrical interpretation
for (\ref{g55}) and (\ref{G55}).

We would like to recollect some of the intriguing features that have
emerged from this study. First, the theory is written more elegantly as
a massless theory in one higher dimension. Second, the interactions are
obtained simply by extending the string to a band. Written in terms
of bands it appears just like a free theory. Both the above features
are reminiscent of M-theory \cite{PT}. The third and probably 
most intriguing
feature is the form of the gauge transformation. It looks like
a local rescaling of the generalized momenta. This is a space-time
scale transformation. The global version of this is of course
just the usual renormalization group (in space-time). The local
(i.e. local along the string)
version of this seems to be a (part of the) gauge group for the string.
This was of course one of the original motivations for this approach
\cite{BS1}.

\appendix

\renewcommand{\thesection}{\Alph{section}}
\renewcommand{\theequation}{\thesection.\arabic{equation}}

\section{Appendix: Basic Definitions}
\label{appena}
\setcounter{equation}{0}

The basic loop variable is 
\[
\lpp = \gvk
\]
where $\al (t)$ is an einbein and $k(t)$ is a distributed momentum.
They have mode expansions: 
\[ \al (t) = \sum _{n \ge 0}\aln t^{-n}\]
\[ k(t) = \sum _{n\ge 0}\kn t^{-n}\]

$Y_n$ is defined by 
\[Y_{n} = \frac{\partial Y}{\partial x   _{n}}. \]
 where   
\[
Y = 
\sum _{n}\aln \frac{\partial^{n}X}{(n-1)!} \equiv
\sum _{n} \alpha _{n} \tilde{Y}_{n} .
\]
and $\xn$ are defined by 
 \[
\sum _{n\ge 0} \aln t^{-n}= e^{\sum _{n\ge 0}\xn t^{-n}}.
\]
The $\aln$ satisfy
\[
{\p \aln \over \p \xm }= \al _{n-m} \; .
\]
Using the above definitions one can easily show that
\[
\dsnm Y = \dsq Y .
\]

The definitions of $\tS$ and $\tG$ are as follows. Define,
(using the notation $z_{i} = z(\sigma _{i})$)
\be     \label{3.10}
D_{z_{1}} = D_{z(\sigma _{1})} \equiv 1+\ai 
(\sigma _{1})\frac{\partial}
{\partial z (\sigma _{1})} + \at \frac{\pp}
{\partial z^{2} (\sigma _{1})}+...
\ee
so that
\be    \label{3.11}
Y(z(\sigma ))=D_{z(\sigma )}X(z(\sigma ))
\ee
then,
\be     \label{3.12}
\tilde{G}(z_{1},z_{2}) = D_{z_{1}} D_{z_{2}}G(z_{1},z_{2})
\ee
\be     \label{3.13}
\tS(\sigma _{1},\sigma_{2}) =  D_{z_{1}} 
D_{z_{2}}\rho (\sigma _{1},\sigma _{2})
\ee
where 
\be     \label{3.14}
\rho (\sigma _{1} , \sigma _{2})=
\frac{\mu (z(\sigma _{1}))-\mu (z(\sigma _{2}))}
{z(\sigma _{1})-z(\sigma _{2})}
\ee
is the generalization of the usual Liouville mode $\rho (\sigma )$
which is equal to $\frac {d \mu }{d z}$.  The $\tilde {\Sigma}$
dependence in the loop variable is obtained by the following step:
\be     \label{3.15}
e^{:\frac{1}{2}\int du \mu (u) [\partial _{z} X(z+u)]^{2}:}
e^{ik_{n}\frac{\partial}{\partial x_{n}} D_{z_{1}}X}
e^{ip_{m}\frac{\partial}{\partial x_{m}}D_{z_{2}}X}
\ee
defines the action of the Virasoro generators on the two sets of 
vertex operators.
\be     \label{3.16}
= e^{ik_{n}.p_{m}\partial _{x_{n}}\partial _{y_{m}}D_{z_{1}}D_{z_{2}}
\oint du\frac{\mu (u)}{z_{1}-z_{2}}
[\frac{1}{z_{1}-u} -\frac{1}{z_{2}-u}]}
\ee
\be     \label{3.18}
=e^{ik_{n}.p_{m}\partial _{x_{n}}\partial _{y_{m}} \tilde{\Sigma}}
\ee     
This expression is only valid to lowest order in $\mu $
which is all we need here.\footnote{The exact expression is given in
\cite{BSV}}

\section{Appendix: Cubic Term in Vector Particle Equation}
\label{appena}
\setcounter{equation}{0}

We start with loop variable (\ref{VLV}),

\[
e^{\int d \si \int d\st \ko (\si ).\ko (\st )[\tG (\si ,\st ) + 
\tS (\si , \st )]} 
\] 
\[
e^ {\int d \si \int d\st \{ \ki (\si ).\ko (\st ) \dsis
 [\tG (\si ,\st ) + \tS (\si , \st )] +\si \leftrightarrow \st \}}
\]
\be   
e^{i\int d\s [\ko (\s )\tY _0(z ) + K_1 (\s ,z-z (\s ))\tY _1 (z)]}
\ee

and keep terms involving three $\ki $'s. This will produce terms
of the form $A.pA.q A$. We can compare these with standard results
from say \cite{GSW}. If they match, then by gauge invariance, other
terms are bound to agree as well.

\[
e^{\int d \si \int d\st \ko (\s _5 ).\ko (\s _6 )[\tG (\s _5 ,\s _6 ) + 
\tS (\s _5 , \s _6 )]} 
\] 
\[{1\over 2!}
\int d \si \int d\st \{ [\ki (\si ).\ko (\st ) \dsis
 \tG (\si ,\st )] +\si \leftrightarrow \st \}
\]
\[
\int d \s _3\int d\s _4 \{ [\ki (\s _3 ).\ko (\s _4 )
 {\p \over \p x_1 (\s _3)}
 \tG (\s _3,\s _4 ) ] +\si \leftrightarrow \st \}
\]
\be
iK_1 (\s , -z )^\mu \tY_1 ^\mu (0)e^{i\ko \tY_0(0)}
\ee

Let us call the three distinct locations $\s _I,\s _{II},\s _{III}$.

Let
\[
k_0(\s _I)=p \,; \, z(\s _I)=z_1
\]
\[
\ko (\s _{II})=q \,; \, z(\s _{II})= z_2
\]
\[
\ko (\s _{III}=k \,; \, z(\s _{III})=z_3
\]

The following is one possible assignment:

\[
\si = \s _I \, ; \, \st = \s _{II} \, or \, \s _{III}
\]
\[
\s _3 = \s _{II} \, ; \, \s _4 = \s _I \, or \, \s _{III}
\]
\[
\s = \s _{III}
\]
$\s _5 $ and $\s _6$ can equal any of them.

When $\s_5 =\s _6$ we need to point split. This gives us
\[
(\eps )^{p^2+q^2+k^2}
(z_1-z_2)^{2p.q}
(z_2-z_3)^{2q.k}
(z_3-z_1)^{2k.p}
\]
\[
{1\over 2!} \{ [{p_1.q\over z_1-z_2}+{p_1.k\over z_1-z_3}]
[{q_1.p\over z_2-z_1}+{q_1.k\over z_2-z_3}]K_1^\mu \tY_1^\mu (0)
e^{i(p+q+k)\tY_0(0)}\}
\]

The other assignments of $\s$'s
 give similar terms which include all premutations
of $p,q,k$.

We are left with integrals over $z_1$ and $z_2$.
(we set $z_3=0$). After changing variables to $z_2'={z_2\over z_1}$
and integrating over $z_2'$, we get the following terms 
(and permutations 
to symmetrize in
$p,q,k$):
\be   
i) \,\,
-\int dz_2' (1-z_2')^{2p.q-2}(z_2')^{2q.k}p_1.qq_1.p =
 -B(2p.q-1,2q.k+1)p_1.qq_1.p 
\ee
\be
ii) \,\, \int dz_2' (1-z_2')^{2p.q-1}(z_2')^{2q.k-1}p_1.qq_1.k =
 -B(2p.q,2q.k)p_1.qq_1.k
\ee
\be
iii) \, \, -\int dz_2' (1-z_2')^{2p.q-1}(z_2')^{2q.k}p_1.kq_1.p =
 -B(2p.q,2q.k+1)p_1.kq_1.p 
\ee
\be
iv) \, \, \int dz_2' (1-z_2')^{2p.q}(z_2')^{2q.k-1}p_1.kq_1.k =
 B(2p.q+1,2q.k)p_1.kq_1.k 
\ee

Using the expansion
\[
B(x,z)=({1\over x}+{1\over z})(1- \zeta (2) xz)+...
\]
\[
={1\over x} + {1\over z} - \zeta (2) (z+x)+...
\]

we see that i) corresponds to a tachyon pole. So we will not  compare
it with the results of \cite{GSW}. The remaining three and 
their permutations
can be compared.
We also use the on-shellness conditions $p^2=q^2=k^2=0$ as well as 
transversality, $p_1.p=q_1.q=k_1.k=0$. One can explicitly check that the
leading pole terms corresponding to vector exchange as well as 
the contact
terms (proportional to $\zeta (2)$ in the above equations) agree with
the corresponding expressions in \cite{GSW}.

\end{document}